\documentclass[showpacs,amsmath,amssymb,twocolumn, 10pt]{revtex4}
\usepackage{graphicx,color}
\usepackage{amsmath}
\usepackage[font=scriptsize]{caption}
\usepackage{subcaption}
\usepackage{float}
\usepackage{color}
\usepackage{amssymb}
\usepackage{bbm}

\usepackage{amsfonts}
\usepackage{bm}

\newcommand{\ud}{\mathrm{d}}

\newcommand{{\Cd}}{{\mathbb{C}^d}}

\newcommand{\tr}{\mathrm{Tr}}

\newcommand{\hh}{\mathcal{H}}

\def\<{\langle}
\def\>{\rangle}

\newtheorem{definition}{Definition}

\newtheorem{Example}{Example}

\newcommand{\beq}{\begin{equation}}
\newcommand{\eeq}{\end{equation}}
\newcommand{\bear}{\begin{eqnarray}}
\newcommand{\ear}{\end{eqnarray}}
\newcommand{\bdm}{\begin{displaymath}}
\newcommand{\edm}{\end{displaymath}}

\newcommand{\sx}{\sigma_1}
\newcommand{\sy}{\sigma_2}

\newcommand{\Las}{\Lambda_t^{(S)}}
\newcommand{\Lae}{\Lambda_t^{(E)}}
\newcommand{\dte}{D_t^{(E)}(\rho_1,\rho_2;\omega)}
\newcommand{\dts}{D_t^{(S)}(\omega;\rho_1,\rho_2)}
\newcommand{\om}{\omega}
\newcommand{\ii}{\mathcal{I}_t(\rho_1,\rho_2,\om)}

\begin{document}
\title{\textbf{Exchange of information between system and environment: facts and myths}
}
\author{Filip A. Wudarski and Francesco Petruccione}
\affiliation{Quantum Research Group, School of Chemistry and Physics,
University of KwaZulu-Natal, Durban 4001, South Africa,
and National Institute for Theoretical Physics (NITheP), KwaZulu-Natal, South Africa}

\begin{abstract}
The exchange of ``information'' between a system and its environment based on the reduced dynamics is investigated. The  association of trace distance with information cannot be stated, because of lack of symmetry between leakage from the system and absorbability by the environment. A measure of loss for the reduced dynamics is established, which may be seen as a deviation from exact unitary dynamics. 
\end{abstract}
\maketitle


\section{Introduction}
Modern quantum technologies cannot neglect effects connected with the inevitable interaction with the surrounding environment \cite{Breuer, Weiss, Alicki}. Several techniques allow us to take into account the influence of the system-environment interaction, such as those based on reduced dynamics \cite{GKS, Lindblad}, stochastic Schr\"odinger equation \cite{partha, barch} or non-Hermitian Hamiltonians \cite{nonH}. The most frequently used one is the reduced dynamics approach that leads to a dynamical map of the system of  interest. The intensive study of the dynamical map recently revealed a plethora of features of open quantum systems dynamics, from which it is worth emphasising the Markovian and non-Markovian behaviour \cite{rev1, rev2}. Markovianity is based on the property of either $P$-divisibility \cite{WissBLP} or $CP$-divisibility \cite{RHP, riv} of the corresponding dynamical map. With $P$-divisibility, we can associate a physical interpretation \cite{BLP}, namely, the distinguishability of quantum states during the evolution. For Markovian behaviour, we monotonically lose ``information'', which flows away to the environment, while for non-Markovian behaviour this ``information'' may come back from the environment to the system. This is called {\it backflow of the information}.

Our work shows that we cannot speak of the ``flow of information'' as a physical quantity. This was proposed in \cite{laura} without deeper insight. According to \cite{rev2, WissBLP, BLP, BR, Laine} the ``information'' that flows from the open system is transferred into environment or system-environment correlations. We demonstrate that ``information'' absorbed by the environment might be greater then the one that has flown from the system (we also refer to it as a leakage of ``information''). Therefore, we claim, that this quantity is ill-defined in a physical way.

During the procedure of the reduction (that is necessary to obtain a quantum dynamical map), we lose some information about the influence of the environment. In this paper, we propose a  quantity that can measure the discrepancies between exact (unitary) dynamics and the reduced one (i.e. dynamics of marginals). It is based on the relative entropy and as we show it cannot be expressed in terms of the trace distance. 

In the next section, we present basics concepts concerning reduced dynamics and distinguishability of the states.  We show the lack of symmetry between reduced dynamics and introduce the relative entropy between exact and reduced dynamics.  In Section III, we consider simple examples that clarify our approach. In the last section, we conclude our work and pose some open questions. 

\section{Reduced dynamics and the information about evolution}
Reduced dynamics $\Lae$ of the system-environment evolution is defined as follows
\begin{equation}
\rho(t)=\Lae\big(\rho(0);\om(0)\big)=\tr_E\Big(U_t \rho(0)\otimes \om(0) U_t^\dag\Big),
\end{equation}
where $\rho(t)\in S(\hh_S)$ is a state of the system living on Hilbert space $\hh_S$, $\om(t)\in S(\hh_E)$ is a state of the environment and $U_t$ is a unitary operation corresponding to the total system-environment evolution. To deal with a dynamical map, one has to consider the initial system-environment state as a product state. Usually, the state of the environment is omitted in the dynamical map. However, we keep it as a parametric dependence on our evolution. According to the Stinespring Theorem \cite{stine} a dynamical map is completely positive (CP), trace preserving (TP) and at the initial value it is the identity map, i.e. $\Lambda_0^{(E)}=\mathbbm{1}$. Therefore, $\Lae$ defines a legitimate physical process.

We may use the same procedure to construct the reduced dynamics of the environment
\begin{equation}
\om(t)=\Las\big(\om(0);\rho(0)\big)=\tr_S\Big(U_t \rho(0)\otimes \omega(0) U_t^\dag\Big).
\end{equation}
In contrast to the previous approach, we traced out the system and left only the environment. This dynamical map is also CPTP with the identity map at the beginning of the evolution.  

\subsection{Distinguishability of the states}
Let us recall a definition of Markovian dynamics based on the distinguishability of the states \cite{BLP}, which was recently proven to be equivalent to the $P$-divisibility \cite{WissBLP} (in this work we will limit ourselves to a standard definition from \cite{BLP}). 

First let us introduce the notion of trace distance. 
\begin{definition}
A trace distance between two states $\rho_1$ and $\rho_2$ is defined as
\begin{equation}
D(\rho_1,\rho_2)=\frac{1}{2}\|\rho_1-\rho_2\|_1,
\end{equation}
where $\|A\|_1=\tr\sqrt{AA^\dag}$ is  the trace norm.
\end{definition}
It is a measure of distinguishability of the state. Due to the fact, that trace distance is contractive under the action of CPTP, it may be used as a witness of non-Markovianity.
\begin{definition}[Markovian dynamics]
A dynamical map $\Lambda_t$ is said to be Markovian if for all pairs of quantum states $\rho_1$ and $\rho_2$  and all times $t\ge0$ it satisfies
\begin{equation}\label{mark}
\frac{\ud}{\ud t}D\Big(\Lambda_t(\rho_1),\Lambda_t(\rho_2)\Big)\le0.
\end{equation}
If condition (\ref{mark}) is violated, then the dynamics is called  non-Markovian. 
\end{definition}
 According to the above definition, the trace distance is monotonically decreasing for Markovian dynamics. This is interpreted as a leakage of the ``information'' from the system to the environment and hence discrimination of two quantum states is monotonically decreasing. Therefore, one should expect that the information outside the system should be contained in one or both of the following:
 \begin{enumerate}
 \item environment,
 \item system-environment correlations.
 \end{enumerate}
 Moreover, ``information'' that leaked from the system supposed to be equal to ``information'' absorbed by the environment or available in correlations.
 
Let us introduce the time evolution of trace distance as
 \begin{equation}\label{dist}
 D_t^{(E)}(\rho_1,\rho_2;\om)=\frac{1}{2}\|\Lae(\rho_1;\om)-\Lae(\rho_2;\om)\|_1.
 \end{equation}
In a similar way we may define the trace distance for the reduced dynamics of the environment as
\begin{equation}\label{distenv}
D_t^{(S)}(\om;\rho_1,\rho_2)=\frac{1}{2}\|\Las(\om;\rho_1)-\Las(\om;\rho_2)\|_1.
\end{equation}
This definition is in accordance with (\ref{dist}) and it should represent the absorbability of the ``information'' by the environment. 
In \cite{Laine, rev2, WissBLP} the ``information'' outside the system (i.e. in the environment) is defined as
\begin{equation}\label{ext}
\mathcal{I}_{\mathrm{ext}}(t)=\frac{1}{2}\| U_t\rho_1\otimes\om U_t^\dag-U_t\rho_2\otimes\om U_t^\dag\|_1 -D_t^{(E)}(\rho_1,\rho_2;\om),
\end{equation}
which is bounded by 
{\footnotesize
\begin{eqnarray}
\mathcal{I}_{\mathrm{ext}}(t)\le D\Big(U_t \rho_1\otimes\omega U_t^\dag,\Lae(\rho_1;\om)\otimes\Las(\om;\rho_1)\Big)+\\
D\Big(U_t \rho_2\otimes\omega U_t^\dag,\Lae(\rho_2;\om)\otimes\Las(\om;\rho_2)\Big)+D_t^{(S)}(\om;\rho_1,\rho_2),\nonumber
\end{eqnarray}}where first two term on the right-hand side of this inequality are measure of system-environment correlations, while the last one is the above mentioned absorbability of the environment. Let us observe that, if (\ref{dist}) has the meaning of the ``information'' flow from the system outside (environment and creation of correlations) then (\ref{distenv}) represents ``information'' absorbed by the environment and hence $\mathcal{I}_{\mathrm{ext}}(t)\ge \dts$.  

Therefore, we should expect that 
\begin{equation} \label{const}
D_t^{(E)}(\rho_1,\rho_2;\om)+D_t^{(S)}(\om;\rho_1,\rho_2)\le D_0^{(E)}(\rho_1,\rho_2;\om)\le C,
\end{equation}
$C=\mathrm{const}$ and equality holds when correlations are not created. This would indicate, that the  amount of ``information'' that leaked from the system was less then absorbed one. In general (\ref{const}) does not hold. Thus, we cannot refer to the quantities $D_t^{(E)}$, $D_t^{(S)}$ and even $\mathcal{I}_{\mathrm{ext}}(t)$ as a measures of the ``information'', due to the fact, that they are not conserved. This is the reason, that we referred to ``information'' in quotation marks.  In this work we examine the quantity that  shows the difference between initial value of trace distance and sum of  trace distances of the reduced dynamics
{\footnotesize
\begin{equation}\label{ii}
\mathcal{I}_t(\rho_1,\rho_2,\om)=D_0^{(E)}(\rho_1,\rho_2;\om)-\Big(D_t^{(E)}(\rho_1,\rho_2;\om)+D_t^{(S)}(\om;\rho_1,\rho_2)  \Big).
\end{equation}
 }We call this a trace distance difference. Based on the investigation of (\ref{ii}), we may conclude, that a negative value indicates violation of conservation of information. All these effects are connected with information losses during the procedure of reduction. Therefore, we need to know how to quantify them.


\subsection{Relative entropy}
To have a better insight into the amount of information losses during the procedure of reduction, we propose the following quantity
\begin{equation}\label{ent}
S_U^\Lambda(\rho,\om)=S\Big( U_t \rho\otimes \om U_t^\dag\| \Lae(\rho;\om)\otimes \Las(\om;\rho)\Big),
\end{equation}
which is a relative entropy \cite{nilsen} between exact (unitary) dynamics and tensor product of reduced dynamics (dynamics of marginals), where
\begin{equation}
S(\rho\|\sigma)=\tr \rho(\log \rho-\log \sigma).
\end{equation}
One sees that $S_U^\Lambda(\rho,\om)=0$ for $U_t=U_t^{(S)}\otimes U_t^{(E)}$, where $U^{(S/E)}$ is a unitary acting on $\hh_{S/E}$, which is a trivial case. Otherwise, we have the measure of the departure from the exact unitary evolution or, in other words, the measure of losses of applying the reduction procedure. Let us stress that we do not refer to (\ref{ent}) as a measure of non-Markovianity or system-environment correlations, but a measure of informational losses.

Usually, we do not have access to the environment, and we assume some fixed state $\om$. If we minimise (\ref{ent}) over the states $\rho$, then this state ``feels'' $\om$ better than the others, i.e. that information losses is optimal for this kind of dynamics. 


\section{Examples}
Consider a two-dimensional system interacting with a two-dimensional environment. The Hamiltonian of the system-environment is of the following form
\begin{equation}\label{ham}
H=f(t)\sx\otimes\sy,
\end{equation}
where   $\sx, \sy$ are $x$ and $y$ Pauli matrices, respectively.  With this Hamiltonian we may associate the unitary matrix
{\footnotesize
\begin{equation}
U_t=\left(\begin{array}{cccc}\cos\big[ F(t)\big] & 0 & 0 & -\sin\big[ F(t)\big] \\0 & \cos\big[ F(t)\big] & \sin\big[ F(t)\big] & 0 \\0 & -\sin\big[ F(t)\big] & \cos\big[ F(t)\big] & 0 \\\sin\big[ F(t)\big] & 0 & 0 & \cos\big[ F(t)\big]\end{array}\right),
\end{equation}
}where $F(t)=\int_0^t f(\tau)\ud\tau$. Our reduced dynamics are defined as follows
{\footnotesize
\begin{eqnarray}
\Las(\om;\rho)&=&\cos^2\big[F(t)\big] \om+\sin^2\big[F(t)\big]\sy \om \sy+\frac{i}{2}g_1(t)[\om,\sy],\nonumber\\
\Lae(\rho;\om)&=&\cos^2\big[F(t)\big] \rho+\sin^2\big[F(t)\big]\sx \rho \sx+\frac{i}{2}g_2(t)[\rho,\sx],\nonumber
\end{eqnarray}
}where $g_{k}(t)=\sin\big[2 F(t)\big]\tr(\sigma_{k} A_k)$, for $k=1,2$, where $A_1=\rho$ and $A_2=\om$. For sake of simplicity we choose the environmental state as
\begin{equation}
\om(0)=\frac{1}{2}\left(\begin{array}{cc}1 & 1 \\1 & 1\end{array}\right),
\end{equation}
for which $g_2(t)=0$ and we get a random unitary dynamics of a qubit \cite{fi1}. It seems clear that both $\dts $ and $\dte$ are strongly state dependent, and we may have various behaviours of those quantities. Here we present results for two types of states: 
\begin{enumerate}
\item Mutually orthogonal states, that according to \cite{WissBLP} are optimal. These states are of the following form
\begin{equation}
\rho_1=\left(
\begin{array}{cc}
 \sin ^2\left(\frac{\pi }{8}\right) & \sin \left(\frac{\pi }{8}\right) \cos \left(\frac{\pi
   }{8}\right) \\
 \sin \left(\frac{\pi }{8}\right) \cos \left(\frac{\pi }{8}\right) & \cos ^2\left(\frac{\pi
   }{8}\right)\\ 
\end{array}
\right),
\end{equation}
\begin{equation}
\rho_2=\left(
\begin{array}{cc}
 \cos ^2\left(\frac{\pi }{8}\right) & -\sin \left(\frac{\pi }{8}\right) \cos \left(\frac{\pi
   }{8}\right) \\
- \sin \left(\frac{\pi }{8}\right) \cos \left(\frac{\pi }{8}\right) & \sin
   ^2\left(\frac{\pi }{8}\right) \\
\end{array}
\right).
\end{equation}
\item Randomly chosen pair of states (referred as random states):
\begin{equation}
\rho_1=\left(
\begin{array}{cc}
 0.655 & 0.205 -0.225 i \\
 0.205+ 0.225 i & 0.345 \\
\end{array}
\right),
\end{equation}
\begin{equation}
\rho_2=\left(
\begin{array}{cc}
 0.73 & 0.275 -0.045 i \\
 0.275+ 0.045 i & 0.27 \\
\end{array}
\right).
\end{equation}
\end{enumerate}
\begin{Example}[Markovian Semigroup]
A Markovian semigroup is obtained for 
\begin{equation}
f(t)=\frac{2\gamma e^{-2 \gamma t}}{\sqrt{1-e^{-4\gamma t}}},\quad \gamma>0
\end{equation}
with the local-in-time generator of the following form
\begin{equation}
L(\rho)=\gamma(\sx\rho\sx-\rho).
\end{equation}
We will consider here a case, when $\gamma=1$. For orthogonal states we have the trace distances given by the following formulas
\begin{eqnarray}
\dte&=&\sqrt{\frac{1+e^{-4t}}{2}},\\
\dts&=&\sqrt{\frac{1-e^{-4t}}{2}}.\\
\end{eqnarray}
$D^{(E)}_0(\rho_1,\rho_2,\om)=1$. These results as well as $\ii$ are presented in the Fig 1. (b). In the Fig. 1 (a) we examine  the sum of relative entropy of two quantum states $\rho_1$ and $\rho_2$ between exact and reduced dynamics, i.e.
\begin{equation}\label{ents}
\frac{1}{2}\Big(S_U^\Lambda(\rho_1,\om)+S_U^\Lambda(\rho_2,\om)\Big),
\end{equation}
the results for random states are based on numerical calculations, while for orthogonal states we used a computer program.
\begin{figure}[h]
\captionsetup{justification=raggedright
}
\centering
\begin{subfigure}[(a)]{0.43\textwidth}
\includegraphics[width=\textwidth]{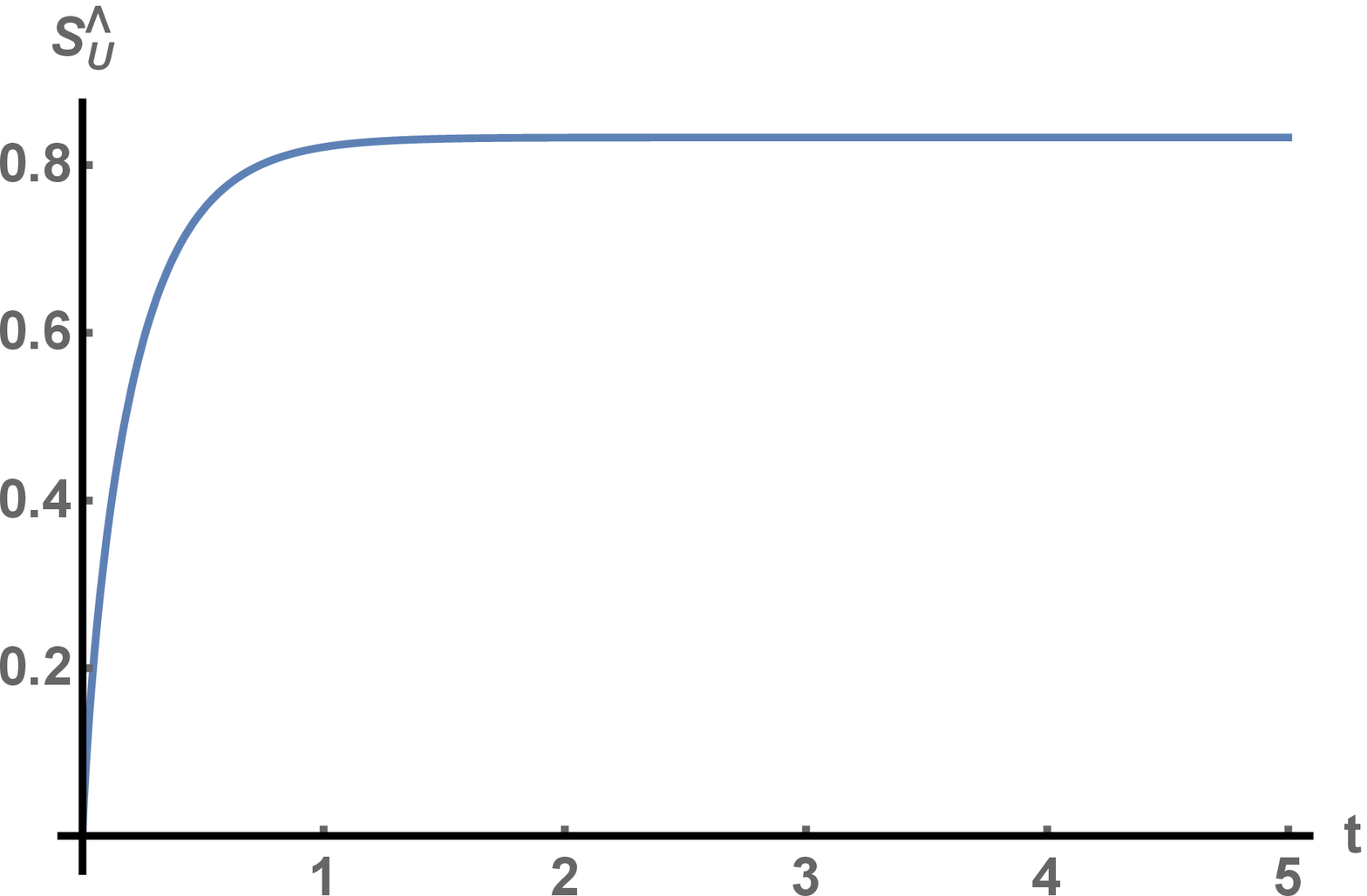}
\caption{}
\end{subfigure}
\begin{subfigure}[(a)]{0.43\textwidth}
\includegraphics[width=\textwidth]{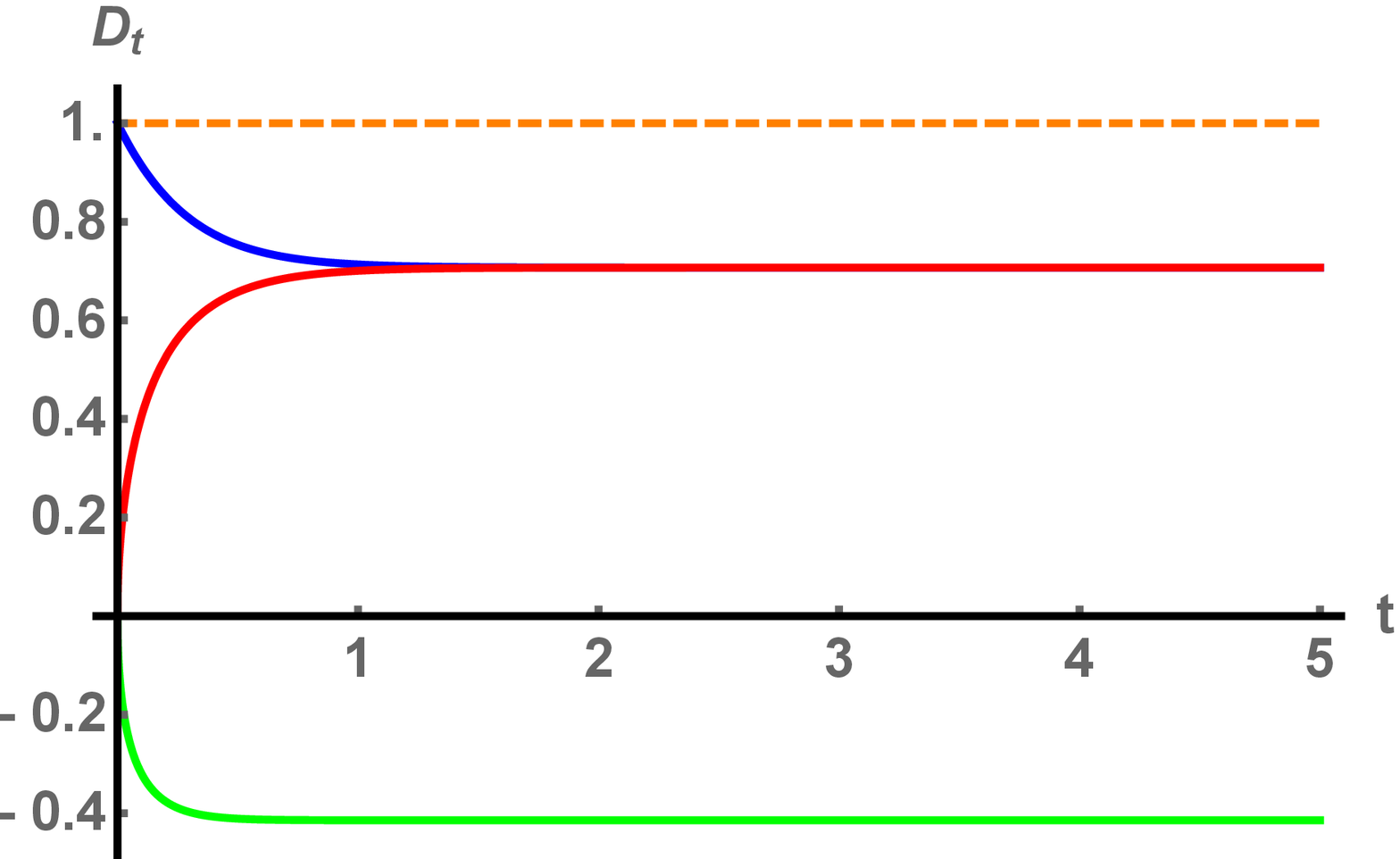}
\caption{}
\end{subfigure}
\caption{Semigroup dynamics for mutually orthogonal states. (a):~The sum of relative entropy between exact unitary dynamics and tensor product of reduced dynamics for both states $\rho_1$ and $\rho_2$. (b):~Trace distances between two orthogonal states. The  blue curve represents $\dte$, the red curve is $\dts$, the dashed line is the initial value of trace distance $D_0^{(E)}(\rho_1,\rho_2;\om)$ and the green one represents the difference of the trace distances $\mathcal{I}_t(\rho_1,\rho_2,\om)$.
}
\end{figure}
For random states one may easily find formulas for trace distances
\begin{eqnarray}
\dte&\approx&0.207\sqrt{0.114+0.886e^{-4t}},\\
\dts&\approx&0.07\sqrt{1-e^{-4t}}.
\end{eqnarray}
For transparency we rounded up our results. Trace distances $\dte$, $\dts$ and $\ii$ are presented in the Fig. 2 (b), while the sum of relative entropies (\ref{ents}) is evaluated numerically (Fig. 2 (a)).
\begin{figure}[h]
\captionsetup{justification=raggedright
}
\centering
\begin{subfigure}[(a)]{0.43\textwidth}
\includegraphics[width=\textwidth]{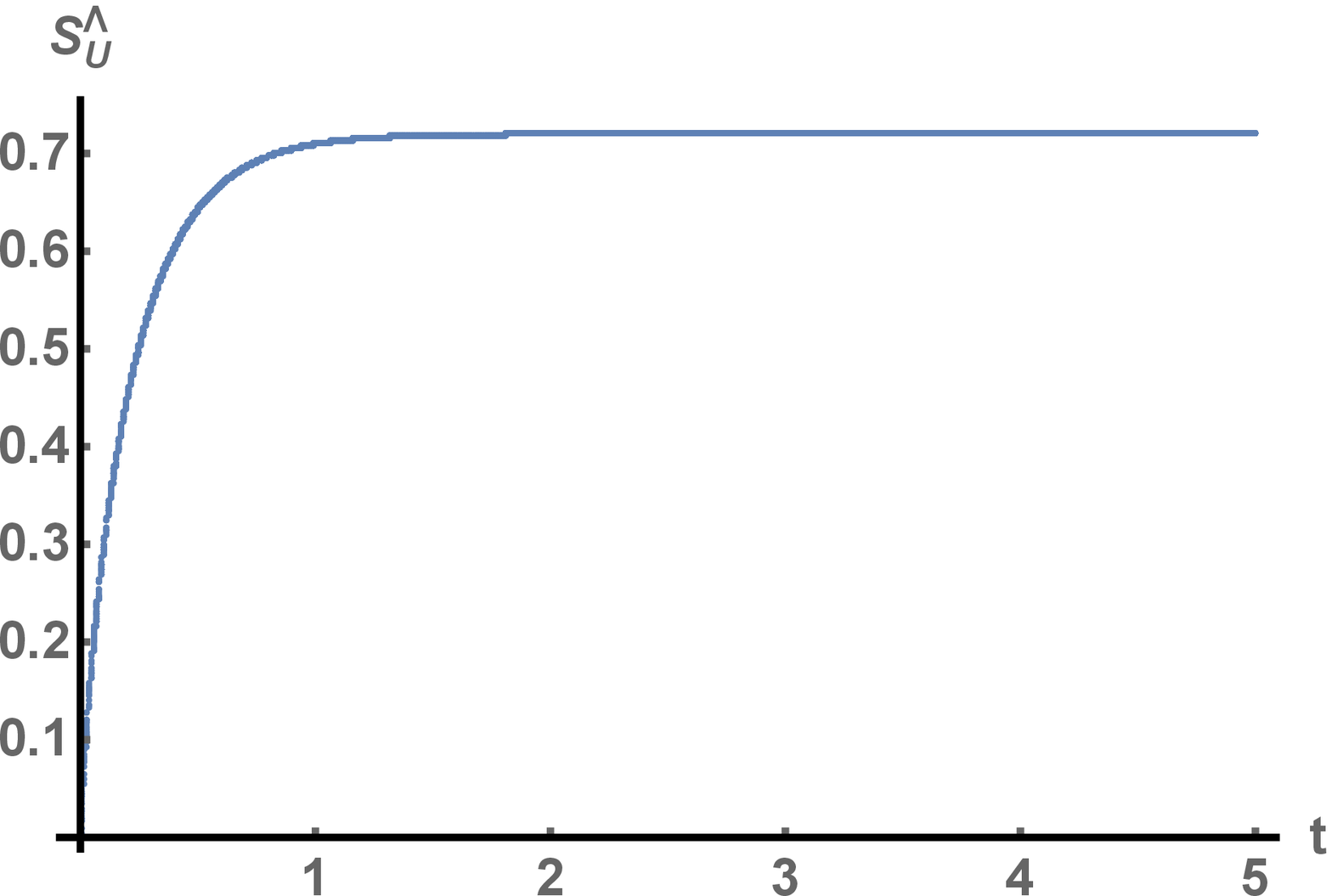}
\caption{}
\end{subfigure}
\begin{subfigure}[(a)]{0.43\textwidth}	
\includegraphics[width=\textwidth]{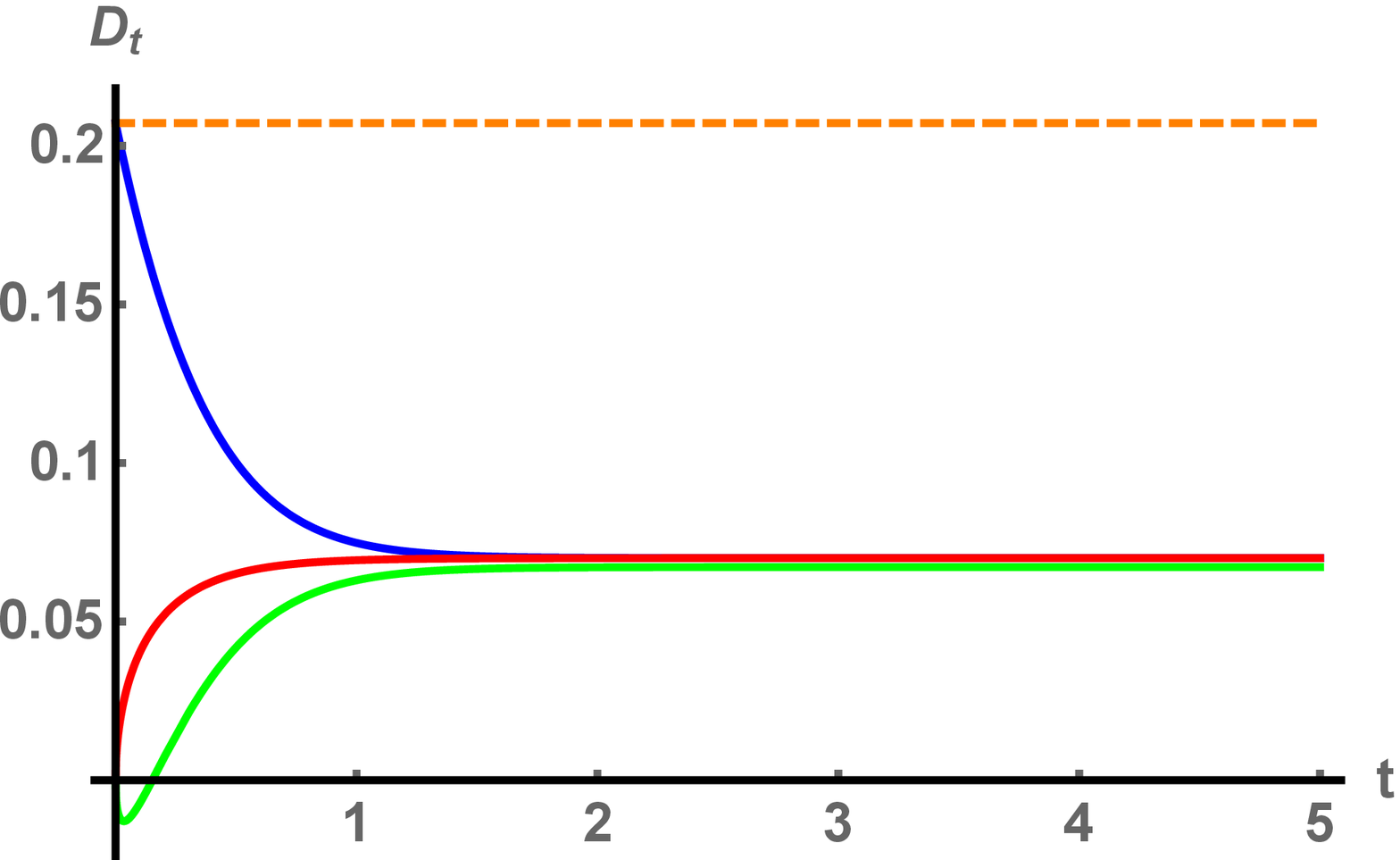}
\caption{}
\end{subfigure}
\caption{Semigroup dynamics for random states. (a):~The sum of relative entropy between exact unitary dynamics and tensor product of reduced dynamics for both states $\rho_1$ and $\rho_2$. (b):~Trace distances between two orthogonal states. The blue curve represents $\dte$, the red curve is $\dts$, the dashed line is the initial value of trace distance $D_0^{(E)}(\rho_1,\rho_2;\om)$ and the green one represents the difference of the trace distances $\mathcal{I}_t(\rho_1,\rho_2,\om)$.
}
\end{figure}
For orthogonal states we have  faster absorbability (Fig. 1 (b) the  red curve) then leakage  (Fig.1 (b) the blue curve) for all $t\ge0$, therefore the trace distance difference is negative (Fig.1 (b) the green curve). This indicates that that the trace distance cannot be understood as a measure of the ``information'', due to lack of conservation law. For random states, we have at the beginning of the evolution faster absorbability then leakage, but as evolution progresses leakage starts to dominate absorbability (Fig. 2 (b) the green curve -- initially negative, then becomes positive). 

Because we are interested in the comparison  between trace distance acting on different states and relative entropy on different states, hence we examine (\ref{ents}). We use the sum instead of the difference, due to the fact, that losses of information cannot compensate each other. Interestingly, the behaviour of relative entropy is regular (in this case even qualitatively the same  for different pair of states (Fig.1 (a) and Fig. 2 (a))), which in general does not need to be true. As the evolution progresses we lose more and more information about the total dynamics. This seems to be clear, due to the fact that Markovian semigroup is a memoryless dynamics and information once lost, cannot be ever recovered. 
\end{Example}

\begin{Example}[Non-Markovian dynamics]
Now consider a non-Markovian dynamics, with a function
\begin{equation}
f(t)=\mu ,
\end{equation}
that leads to a singular generator
\begin{equation}
L_t(\rho)=4\tan(2 \mu t)(\sx\rho\sx-\rho).
\end{equation}
Due to the occurrence of singularity it is better to consider nonlocal-in-time approach 
\begin{equation}
\dot{\rho}_t=\int_0^tK_{t-\tau}(\rho_\tau)\ud \tau,
\end{equation}
where $K_t$ is a memory kernel and may be found by the methods described in \cite{fi2} as
\begin{equation}
K_t(\rho)=2\mu^2 (\sx\rho\sx-\rho).
\end{equation}
Here for simplicity we fix $\mu=1$. The formulas for trace distances may be found as
\begin{eqnarray}
\dte&=&\frac{1}{2}\sqrt{3+\cos(4t)},\\
\dts&=&\frac{1}{2}\sqrt{1-\cos(4t)},
\end{eqnarray}
for orthogonal states. These together with $\ii$ and $D_0^{(E)}=1$ are depicted in the Fig. 3 (b). For random states we have
\begin{eqnarray}
\dte&=&0.207\sqrt{0.557+0.443\cos(4t)},\\
\dts&=&0.07|\sin(2t)|.
\end{eqnarray}
We can see these results in Fig. 4 (b). The relative entropy based on the formula (\ref{ents}) for orthogonal and random states is presented by the plots Fig. 3 (a) and Fig. 4 (a), respectively. 
\begin{figure}[h]
\centering
\captionsetup{justification=raggedright
}
\begin{subfigure}[(a)]{0.43\textwidth}
\includegraphics[width=\textwidth]{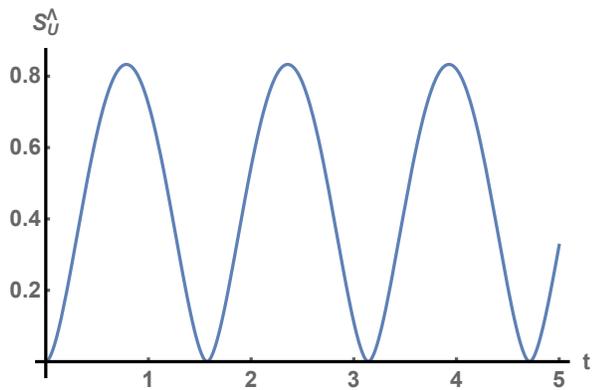}\label{A1}
\caption{}
\end{subfigure}
\begin{subfigure}[(b)]{0.43\textwidth}
\includegraphics[width=\textwidth]{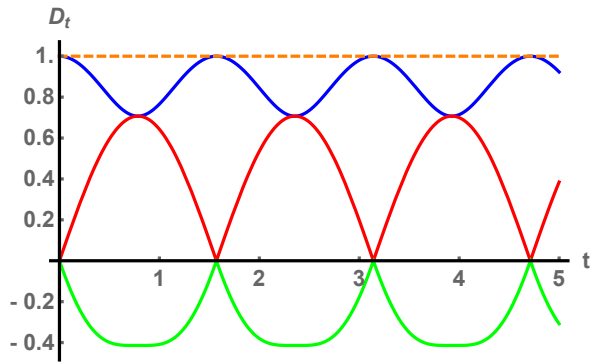}\label{B1}
\caption{}
\end{subfigure}
\caption{Behaviour of non-Markovian dynamics for orthogonal states.  (a):~The sum of relative entropy between exact unitary dynamics and tensor product of reduced dynamics for both states $\rho_1$ and $\rho_2$. (b):~Trace distances between two orthogonal states. The blue curve represents $\dte$, the red curve is $\dts$, the dashed line is the initial value of trace distance $D_0^{(E)}(\rho_1,\rho_2;\om)$ and the green one represents the difference of the trace distances $\mathcal{I}_t(\rho_1,\rho_2,\om)$.}
\end{figure}
\begin{figure}[h]
\captionsetup{justification=raggedright 
}
\centering
\begin{subfigure}[(a)]{0.43\textwidth}
\includegraphics[width=\textwidth]{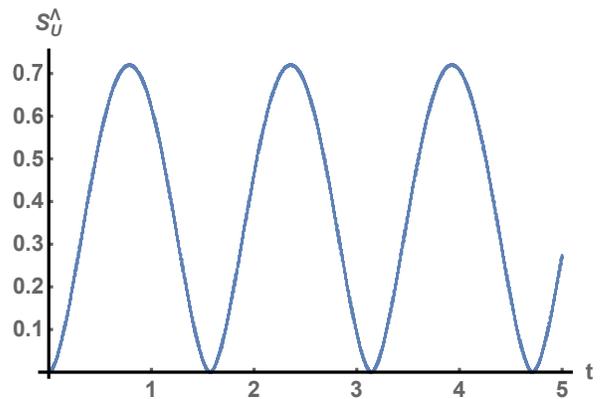}\label{A1}
\caption{}
\end{subfigure}
\begin{subfigure}[(b)]{0.43\textwidth}
\includegraphics[width=\textwidth]{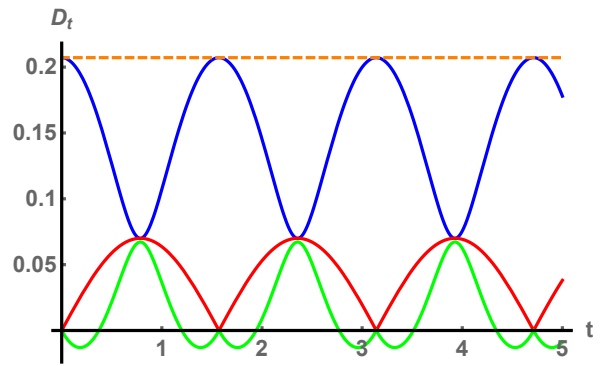}\label{B1}
\caption{}
\end{subfigure}
\caption{Behaviour of non-Markovian dynamics for random states.  (a):~The sum of relative entropy between exact unitary dynamics and tensor product of reduced dynamics for both states $\rho_1$ and $\rho_2$. (b):~Trace distances between two orthogonal states. The  blue curve represents $\dte$, the red curve is $\dts$, the dashed line is the initial value of trace distance $D_0^{(E)}(\rho_1,\rho_2;\om)$ and the green one represents the difference of the trace distances $\mathcal{I}_t(\rho_1,\rho_2,\om)$.}
\end{figure}
This kind of non-Markovian dynamics is periodical in both relative entropy and trace distances. For orthogonal states we have faster absorbability (Fig. 3 (b) the red curve) than leakage (Fig. 3 (b) the blue curve) for all $t>0$, therefore $\mathcal{I}_t(\rho_1,\rho_2,\om)\le0$ (Fig. 3 (b) the green curve).  For random states we have for some periods of time negative trace distance difference (Fig. 4 (b) the green curve), which also indicates faster absorbability, however most of the time, we have faster leakage. Sum of  relative entropies for $\rho_1$ and $\rho_2$ states is in both orthogonal (Fig. 3 (a)) and random states (Fig. 4 (a))  qualitatively the same. 
\end{Example}

\section{Conclusions}
In this work we showed that the trace distance of the reduced dynamics is not symmetrical, i.e. the reduced dynamics over environmental degrees of freedom is different from the reduced dynamics over a system. The difference may be even negative, which indicates that it cannot be understood as a measure of information contained in the systems that is exchanged with the environment during the evolution. Regardless of the absence of informational interpretation the trace distance is a great mathematical tool to decide whether a given map is $P$-divisible or not, i.e. it is a witness of non-Markovianity.  

We introduce a measure of losses of the reduction procedure, which is quantified by the relative entropy of exact unitary dynamics and tensor product of reduced dynamics. This measure is state dependent, and we can minimise it to get the state that is least affected by the procedure of reduction. It would be worth investigating a behaviour of relative entropy for correlated system-environment states, in particular, entangled states. 

All our considerations show that when we neglect the environment by mere reduction we lose some amount of information about our evolution ($S_U^\Lambda$). The procedure of reduction seems to be appropriate if we have no access to (information about) the environment. However, in many physical applications we can prepare our environment in a certain state (for example thermal baths). Therefore, it seems inappropriate to treat a ``partially known'' state as a ``completely unknown'' one and perform the partial trace.  Hence, this leads naturally to the following question:  How may we substitute the procedure of reduction, to have smaller losses in the dynamics?

\end{document}